# PERTURBATIONS OF HELIOSYNCHRONOUS ORBITS IN STABLE KALUZA—KLEIN THEORY

M. Kuassivi
170374 Cotonou, Benin

ABSTRACT
Although the methods and techniques have been greatly improved since the late nineteenth century, the precision on the measurement of the gravitational constant G does not exceed 1 part in $10^3$. Intrinsic variations of G caused by the geomagnetic field may explain the observed dispersion of the laboratory measurements. This involves a coupling between gravitation and electromagnetism (hereafter GE coupling) and a dependance of the effective G constant with latitude and longitude. In this paper I analyse the effects of this coupling in the framework of classical space mechanics by focusing on heliosynchronous orbits. The predictions are found inconsistent with experimental data from the SPOT mission.

Keywords :

1. INTRODUCTION

Improvements in our knowledge of the absolute value of the Newtonian gravitation constant, G, have become slowly over the years. Most other constants of Nature are known to parts per billion. However, G stands mysteriously alone, its history being that of a quantity which remains virtually isolated from the theoretical structure of the rest of physics (Gillies 1997).

The current status of the G terrestrial measurements (precision of 0.1 %) implies either an unknown source of errors or some new physics. In the latter spirit, many theories involve a coupling between gravitation and electromagnetism.

The simplest theories accounting for a GE coupling are those of Kaluza—Klein (Kaluza 1921, Klein 1926). Their use corresponds to the most economical way (the minimum hypothesis) to test the GE coupling. The genuine five—dimensional Kaluza—Klein theories being subject to instabilities (Sokolowski & Golda, 1987), various author have suggested a more acceptable version which includes an additional stabilizing external bulk field.

The minimum hypothesis of a scalar field coupled to gravity implies that Geff varies with the geomagnetic field. Mbelek & Lachièze—Rey (2002) have demonstrated that such theory can be built in order to reconcile the different laboratory measurements of G via an angular dependency of the Geff. Thus, measurements performed near the equator are biased and give low G value.

In this paper, I propose a test of the latter theory by computing the expected on—orbit GE perturbation for heliosynchronous satellites. Heliosynchronous satellites span a large angular domain and are highly sensitive to non spherical potential perturbation.

In section 2, I recall the genuine theory of secular variation and perturbation for heliosynchronous satellite. In section 3, I complete the perturbed gravitational potential including the GE perturbation. In section 4, I discuss the consistency of the theory with respect to the orbital motion of the SPOT satellite.

2. SECULAR VARIATION OF OMEGA IN HELIOSYNCHRONOUS ORBITS

Inclined resonnant orbits exhibit complex motions. Understanding the mechanisms that cause such motions is critical for control strategies. Secular perturbation theory is used to determine isolated resonance characteristics. The most significant perturbations to a nominal Keplerian orbit in the class being considered is the Earth geopotential (tesseral and zonal) effects. The secular and long-period geopotential terms are determined by averaging the geopotential expansion, expressed in terms of classical orbital elements, over the satellite mean anomaly. After averaging the most significant perturbation that remains is associated with the oblateness of the Earth (Ely & Howell, 1997). It has the form:

$$(1) \quad U_{J_2} = GMR_T^2 \cdot \frac{J_2}{r^3} \cdot \frac{3}{2} \left(\frac{z}{r}\right)^2 + f(r)$$

where $R_T$ is the Earth radius, M the Earth mass and $J_2$ is the second degree tesseral harmonic.

This potential is responsible for the apparition of new forces on the satellite. One is central, the other is parallel to the z—axis toward the equator (it tends to reduce the orbit inclination). The latter perturbation yields only secular changes to the angle variables $\omega$ and $\Omega$ (see below Fig. 1).

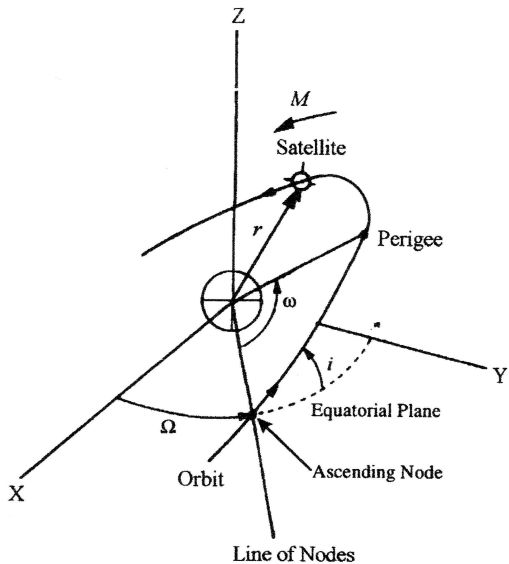

**Fig 1. Earth—centered orbit geometry**

Of particular importance is the secular variation of the ascending node, $\Omega$, which varies linearly with time (Guiziou, R. 2000).

$$(2) \frac{d\Omega}{dt} = -\frac{3 \cdot J_2}{2} \cdot \sqrt{\frac{\mu}{a^7}} \cdot \frac{\cos(i) \cdot R_T^2}{(1-e^2)^2}$$

where I use $\mu = GM$.

This phenomenum is used to control the ascending node of imaging satellites which are located on high inclination resonant orbits. Such satellites are required to take pictures of the Earth with a constant local time (constant light condition). A rapid computation (Guiziou, R. 2000) shows that heliosynchronous satellites must satisfy the following condition:

$$(3) \quad \frac{a}{R_T} = 1.93669 \cdot (-\cos i)^{\frac{2}{7}}$$

where a is the semi—major axis, i the inclination and $R_T$ the earth radius.

3. EFFECTS OF THE GE COUPLING

In the following I will assume an angular dependency of G as described in Mbelek & Lachièze—Rey (2002).

$$(4) \quad G_{\text{eff}} = \frac{G}{\phi}$$

where the scalar field $\phi$ is a function of $\theta,\varphi$ angles and is related to $\hat{g}_{44}$ ($\hat{g}_{44} = -\phi^2$ in the Jordan—Fierz frame). The $\varphi$—angular dependency of G is negligible which gives:

$$(5) \frac{G_{\text{eff}}}{G} \approx 1 + 1.0 \cdot 10^{-3} \cdot \left(\frac{R_T}{r}\right)^4 \cdot \left(\frac{z}{r}\right)^2$$

Where $G = (6.6696 \pm 0.0008) \, 10^{-11}$ m$^3$ kg$^{-1}$ s$^{-2}$ (Mbelek & Lachièze—Rey, 2002). In a straightforward manner, I then replace the G universal constant by the Geff function in order to derive the perturbed potential $U_p$ at first order:

$$(6)\; U_p = U_{J_2} - \frac{4.01 \cdot 10^{11}}{r} \cdot \left(\frac{R_T}{r}\right)^4 \cdot \left(\frac{z}{r}\right)^2$$

The fact that the GE perturbation term is negative has two important consequences. First, this term may eventually cancel out the $J_2$ perturbation and, thus, make heliosynchronicity impossible. Secondly, one can easily show that the additional z—force acting on the satellite points toward the poles. This force can be interpreted as an additional attraction toward the pole due to greater G value (equivalent to an additional mass at the poles). Hence, the GE perturbation is opposed to the oblatness of the Earth.

4. RESULTS AND DISCUSSION

The perturbed gravitational potential is computed for the SPOT satellite (semi—major axis a = 7200 km and inclination of 98.7°). These orbital parameters fit perfectly the expected secular variation (eq. 3) in Keplerian motion and any new perturbation must be several order of magnitude smaller than the $J_2$ term.

However, a rapid computation at the poles ( with z≈r ) shows that $U_{GE} / U_{kepler} = 6.3 \; 10^{-4} \gg U_{J2} / U_{kepler}$ (1.3 $10^{-8}$). The result shows that if the GE perturbation existed it would dramatically change the strategy control of the SPOT satellite.

Note that the $J_2$ term is currently measured from in—orbit motion. As thus, one can argue that any unseen perturbation is already included in the classical computation. However, the amplitude and the sign of the GE perturbation makes it obviously inconsistent with the data.

It is beyond the scope of the paper to discuss the intrinsic variability of the G constant. What is pointed out here is the necessity of reconciling theory and facts using genuine space mechanics.

REFERENCES

Ely, Todd A. & Howell, Kathleen C. 1997, Dynamical Systems, 12, 243

Gillies, G.T. 1997, Rep. Prog. Phys. 60, 151 and references therein

Guiziou, Robert 2000, lecture notes on space mechanic from Université Aix—Marseille I (http://icb.u-bourgogne.fr/uni--versitysurf/ressources-mecanique.html)

Kaluza, Th. 1921, Sitzungsberichte der K. Preussischen Akademie der Wissenshaft Zu Berlin, 966

Klein, O. 1926, Zeitschrift für Physik, 37, 895

Mbelek, J.P. & Lachièze—Rey, M. 2002, GrCo, 8, 331

Sokolowski, M. & Golda, A. 1987, Phys. Lett. B, 83, 349